\documentclass[12pt]{article}
\usepackage{amsmath}
\usepackage{graphicx,psfrag,epsf}
\usepackage{enumerate}
\usepackage{natbib}
\usepackage{url} 

\usepackage{graphicx}
\usepackage{slashbox}
\usepackage{amssymb}
\newtheorem{theorem}{Theorem}
\newtheorem{remark}{Remark}

\newcommand{\blind}{0}

\date{}
\begin{document}

\bibliographystyle{ieeetr} 
\bibliography{refs}

\begin{thebibliography}{50} 
%
\bibitem[Buu \textit{et al.}(2011)]{buu11} Buu A., Joohnson N.J. , Li R. and Tan X.. New variable selection methods for zero-inflated count data with applications to the substance abuse field. Statistics in Medicine, No 30: 2326-2340, 2011.
\bibitem[Chatterjee \textit{et al.}(2018)]{chat18} Chatterjee S., Chowdhury S., Mallick H. , Banerjee P. and Garai B. Group regularization for zero-inflated negative binomial regression models with an application to healthcare demand in Germany.  Statistics in Medicine, No 37: 3012-3026,  2018.
\bibitem[Czado \textit{et al.}(2007)]{czado07} Czado C., Erharddt V. and Wagner S.  Zero-inflated generalized Poisson models with regression effects on the mean, dispersion and zero-inflation level applied to patent outsourcing rates. Statistical Modelling, No 7: 125-153, 2007.
\bibitem[Chin-Shang and Minggen(2022)]{chin22} Chin-Shang L. and Minggen L. Semiparametric zero-inflated Bernoulli regression with applications. Journal of Applied Statistics. Vol. 49, No. 11, 2845–2869, https://doi.org/10.1080/02664763.2021.1925228, 2022.
\bibitem[Diop \textit{et al. }(2011)]{ddd11} Diop A., Diop A. and Dupuy J.F.  Maximum likelihood with a cure fraction, Electron. J. Stat. 5, pp. 460–483, 2011
\bibitem[Diop \textit{et al.}(2016)]{ddd16} Diop A., Diop A.  and Dupuy J.-F.  Simulation-based inference in a zero-inflated bernoulli regression model. Communications in Statistics - Simulation and Computation, 104(67) :45(10) :3597–3614, 2016.
\bibitem[Deb and Trivedi(1997)]{deb97} Deb P. and Trivedi P.K. Demand for medical care by the elderly: a finite mixture approach. Journal of Applied Econometrics, Vol. 12, No. 3, Special Issue: Econometric Models of Event Counts (May - Jun., 1997), pp. 313-336. 1997.
\bibitem[Dupuy(2017)]{dup17} Dupuy, J.F. Inference in a generalized endpoint-inflated binomial regression model. Statistics. 51(4) :888–903, 2017.
\bibitem[Eicker(1966)]{eicker66} Eicker F. The Annals of Mathematical Statistics. 37(6), 1825-1828,  1966.
\bibitem[Fong and Yip (1995)]{fong95} Fong D.Y.T.. and Yip P.S.F. A note on information loss in analysing a mixture model of count data, 1995.
\bibitem[Friedman \textit{et al.}(2010)]{fried10} Friedman J., Hastie T. and Tibshirani R. Regularization Paths for Generalized Linear Models via Coordinate Descent. J Stat Softw [Internet];33(1):1–20. 2010.
\bibitem[Greenland (2000)]{green2000} Greenland S. Principles of multilevel modelling. in. j. Epidemiology, 29 :158167, 2000.
\bibitem[James \textit{et al.}(2013)]{james13} James G., Witten D., Hastie T. and Tibshirani R. An Introduction to Statistical Learning [Internet]. Springer Texts in Statistics. 2013. 
\bibitem[Derksen and Keselman(1992)]{derk92} Derksen S. and Keselman H.J. Backward, forward and stepwise automated subset selection algorithms: Frequency of obtaining authentic and noise variables. Br J Math Stat Psychol;45(2):265–82. 1992.
\bibitem[Zou(2006)]{zou06} Zou H. The adaptive lasso and its oracle properties. Journal of the
American Statistical Association, 101 :14181429, 2006.
\bibitem[Zou (2005)]{zou05} Zou H. Regularization and variable selection via the elastic net. Journal of Royal Statistical Society, Series B, 67(2) :301320, 2005.
\bibitem[Henningsen and Toomet (2011).]{hen11} Henningsen A. and Toomet O. maxlik : A package for maximum likelihood estimation in R. Computational Statistics. 117, 2011.
\bibitem[Hilbe(2009)]{hilbe09} Hilbe J. M. Logistic regression models. PhD thesis, Chapman and
Hall : Boca Raton, 2009.
\bibitem[Gourieroux and Monfort(1981)]{gm81} Gouriéroux C., Monfort A., 1981. Asymptotic properties of the maximum likelihood estimator in dichotomous logit models. Journal of Econometrics 17: 83–97.

\bibitem[Guyon(2001)]{g01} Guyon X., 2001. Statistique et économétrie - Du modèle linéaire aux modèles non-linéaires. Ellipses Marketing, 2001.
\bibitem[Hoerl (1970)]{hoerl70} Hoerl A. E. and Kennard R. W.. Ridge regression : Applications to non-
orthogonal problems. Technometrics, 12(1) :6982., 1970.
\bibitem[Lo \textit{et al.} (2022)]{lo22} Lo F., Ba, D.B., Diop A., 2022. Maximum likelihood estimation in the generalized extreme value regression model for binary data. Gulf Journal of Mathematics. ISSN: 2309-4966. Vol 12, Issue 2, pp: 49-56.
\bibitem[Preisser \textit{et al.} (2012)]{prei12} Preisser J. S. , Stamm G.W. , Long D. L. and Kincade, M. E. Review and recommendations for zero-inflated count regression modeling of dental caries indices in epidemiological studies. Caries Research, 54(4) :413–423, 2012.
\bibitem[Jiang(2022)]{jiang22} Jiang J. Large sample techniques for statistics. Springer. 2022
\bibitem[Mullahy (1997)]{} Mullahy, J. Heterogeneity, excess zeros, and the structure of count data models. Journal of Applied Econometrics, page 337–350, 1997.
\bibitem[Ridout \textit{et al.}(1998)]{ridout98} Ridout M.  Dem\'etrio C.G.B.and Hinde J. Models for count data with many zeros.  International Biometric Conference, Cape Town, December, 1998.
\bibitem[Tibshirani(1996)]{tibs96} Tibshirani R. Regression shrinkage and selection via the lasso. Journal of Royal Statistical Society B, 58(1), 1996.
\bibitem[Cessie and Van Houwelingen(1992)]{ces92}  Cessie S.LL.  and Van Houwelingen J.C.  Ridge Estimators in Logistic Regression.  Journal of the Royal Statistical Society. Series C (Applied Statistics), Vol. 41, No. 1, pp. 191-201, 1992.
\bibitem[Schwarz (1978)]{schw78} Schwarz G. Estimating the dimension of a model. Ann. Statist. 6(2) :461–464, 1978.
\bibitem[Tu(2002)]{tu02} Tu W. Zero-inflated data. Encyclopedia of Environmetrics, (4) :2387–2391,  2002.
\bibitem[Foutz(1977)]{foutz77} Foutz R.V. On the unique consistent solution to the likelihood equations. Journal of the American Statistical Association, 72 :147–148, 1977.
\bibitem[Wang \textit{et al.}(2015)]{wang15} Wang Z., Shuangge M. and Wang C.Y. Variable selection for zero-inflated and overdispersed data with application to health care demand in Germany. Biom J., Jun 8;57(5):867–884. doi: 10.1002/bimj.201400143, 2015.
%
\end{thebibliography}


\def\spacingset#1{\renewcommand{\baselinestretch}%
{#1}\small\normalsize} \spacingset{1}


  \title{Regularized zero-inflated Bernoulli regression model}
  \author{\textbf{Mouhamed Ndoye}$^{(1)}$\footnote{E-mail: ndoyethiat@gmail.com} ~ and \textbf{Aba Diop}$^{(1)}$\footnote{Corresponding author: e-mail: aba.diop@uadb.edu.sn} ~\\~\\
 $^{(1)}$ \textit{Equipe de Recherche en Statistique et Modèles Aléatoires} \\
 \textit{D\'epartement de Math\'ematiques}\\
  \textit{Université Alioune Diop, Bambey, Sénégal} 
  }
  
\maketitle
\if0\blind

\begin{abstract}
Logistic regression model is widely used in many studies to investigate the relationship between a binary response variable $Y$ and a set of potential predictors $X_1,\ldots, X_p$ (for example: $Y = 1$ if the outcome occurred and $Y = 0$ otherwise). One problem arising then is that, a proportion of the study subjects cannot experience the outcome of interest. This leads to an excessive presence of zeros in the study sample. This article is interested in estimating parameters of the zero-inflated Bernouilli regression model in a high-dimensional setting, i.e. with a large number of regressors.  We use particulary Ridge regression and the Lasso which are typically achieved by constraining the weights of the model. and are useful when the number of predictors is much bigger than the number of observations.  We establish the existency, consistency and asymptotic normality of the proposed regularized estimator. Then, we conduct a simulation study to investigate its finite-sample behavior, and application to real data.
\end{abstract}

\noindent%
{\it Keywords:} Regression model, Rgularization, Lasso, Ridge, Elastic net, Model selection, Information criterion.

\spacingset{1.45} 
\section{Introduction}
\label{sec:intro}
Bernoulli regression model has been extensively used to study the relationship between a binary outcome variable and a set of covariates in many disciplines, including, e.g. biomedical studies, epidemiological studies, and social studies (\citep{hilbe09}).  However, many practical data such as healthcare utilization data often contain large numbers of zeros (i.e. there is a large number of non-users of the corresponding healthcare service over the study period), \citep{czado07}, \citep{ddd11} and \citep{chat18}. When there are more zeros than expected under a standard Bernoulli regression model, the data are said to be zero-inflated \citep{tu02}.
\par The zero inflation logistic regression model is a statistical method used to model data with a binary response variable where the probability of observing a zero is higher than would be expected in a standard logistic regression model (\citep{ddd16} and \citep{chin22}). \citep{ddd16} proposed a parametric ZIB regression model for this type of binary outcome data with linear predictors via logit links and applied it to analyze an example of data from a study of dengue fever to investigate the effects of covariates on dengue infection. \citep{ddd11} established the asymptotic properties of the maximum likelihood estimators of the parameters of a parametric ZIB regression model.
\par Among the many potential risk factors, researchers and practitioners are often interested in identifying a small subset. A parsimonious model often offers a better interpretation of the results, particularly in the health field (\citep{wang15}). In the statistical literature, variable selection is one of the most active areas of research. Penalized regression with a variety of penalty functions is a popular choice. For more details, see forexample \citep{tibs96},  \citep{zou05, zou06}. However, there is a lack of methods in selecting variables in a zero-inflated and overdispersed count data framework, and recent research has focused on penalized ZIP models (\citep{czado07}, \citep{buu11} and \citep{prei12}). Penalized regression methods are a useful theoretical approach for both developing predictive models and selecting key indicators within an often substantially larger pool of available indicators.
\par Additionally, regularisation aids in balancing the accuracy and generalisability of predictive models in terms of complexity. Complexity often refers to the number of indicators in the final model. For several penalised regression procedures, the regularisation process results in the coefficients of unimportant variables being shrunk to zero.\\
In the current paper, we develop three regularization methods for the zero-inflated Bernoulli regression model. Specificaly, three penalised regression methods which conduct automatic feature commonly compared in the literature and discussed in standard statistical texts (\cite{derk92},  \cite{james13}), are the least absolute shrinkage and selection operator (LASSO) (see \cite{tibs96}), the ridge regression (see \cite{zou06}) and the elastic-net (see \cite{zou05}). The LASSO applies the $L_1$ penalty (constraint based on the sum of the absolute value of regression coefficients), which shrinks coefficients equally and enables automatic feature selection. However, in situations with highly correlated indicators the LASSO tends to select one and ignore the others (\cite{fried10}). The ridge regression and elastic-net are extensions on the LASSO, both of which incorporate the $L_2$ penalty (\cite{ces92}).\\
In addition to selecting between alternative penalised regression methods, analysts need to tune (i.e., select) the parameter lambda ($\lambda$), which controls the strength of the penalty terms. This is most commonly done via the data driven process of $k$-fold cross-validation \cite{tibs96}.
\par The rest of the paper is organized as follows. Section \ref{sec:models} introduces a penalized zero-inflated Bernoulli (ZIB) regression model, and presents statistical properties and estimation methods. In Section \ref{sec:asymptotic} we establish the asymptotic propreties of the proposed estimator under some regularity conditions. A simulation study is conducted in Section \ref{sec:simul} to evaluate the proposed methods and to investigate the finite-sample behavior of the proposed estimators. In Section \ref{sec:application}, the proposed methods are applied to a real data set. Discussion and perspectives are provided in Section \ref{sec:disc}.

\section{Models}
\label{sec:models}
In this section, we present the zero-inflated Bernnoulli regression model and the regularization methods used (LASSO, ridge and elastic-net).
\subsection{Zero-inflated Bernnoulli regression model (ZIBerRM)}
Let $(Y_1, \mathbf{X}_{1},\mathbf{Z}_{1}),\ldots,(Y_n,\mathbf{Z}_{n},\mathbf{X}_{n})$ be independant and identically distributed copies of the random vector $(Y,\mathbf X,\mathbf Z)$ defined on the probability space $(\Omega,\mathcal{A},\mathbb{P})$. For every individual $i=1, \ldots,n$, $Y_i$ is a binary response variable. \\
In biomedical studies, for example, the binary response variable $Y$ may be the infection status with respect to some disease (that is, $Y_i=1$ if the $i$-th individual is infected, and $Y_i=0$ otherwise). The existence of a proportion of the study subjects (the so-called cured individuals, as opposed to the susceptibles) who cannot experience the outcome of interest can cause to an excessive presence of zeros ($Y_i=0$). It is usually unknown who are the cured and the susceptible subjects, unless the outcome of interest has been observed. Zero-inflation can cause overdispersion (but accounting for zero-inflation does not necessarily remove overdispersion).\\
Let $\mathbf X_i=(1, X_{i2}, \ldots, X_{ip})'$ and $\mathbf Z_i=(1, Z_{i2}, \ldots, Z_{iq})'$ be random vectors of predictors or covariates (both categorical and continuous predictors are allowed). We shall assume in the following that the $\mathbf X_i$'s are related to the status of interest, while the $\mathbf Z_i$'s are related to the zero-inflation. $\mathbf X_i$ and $\mathbf Z_i$ are allowed to share some components.\\
The zero-inflated Bernnoulli regression model for zero-inflated data which we will present here is the following mixture model:
\begin{equation}\label{m1}
		Y \sim ZIBerRM(1,p,\pi):=\pi\mathcal{B}(0)+(1-\pi)\mathcal{B}(p)
	\end{equation}
Equivalently, we have:
\begin{equation}
Y_{i} \sim 
		\begin{cases}
			0 & \quad \text{with probability} \quad \pi_{i} \\
			\mathcal{B}(1,p_{i}) & \quad \text{with probability} \quad 1-\pi_{i}
		\end{cases}
\end{equation}
	\begin{equation}\label{m2}
		\mathbb{P}(Y_{i}=y)=
		\begin{cases}
			\pi_{i}+(1-\pi_{i})(1-p_{i}) & \text{si $y=0$}              \\
			(1-\pi_{i})p_{i} & \text{si $y=1$}
		\end{cases}
	\end{equation} 
where the parameters $\pi=(\pi_1,\ldots, \pi_n)'$ and $p_i=(p_1,\ldots, p_n)'$ are modeled through \textit{logit} link generalized linear models as:
\begin{equation}\label{m3}
p_i= \frac{e^{\beta'X_{i}}}{1+e^{\beta'X_{i}}}
\quad\quad\text{and}\quad\quad
\pi_i  =\frac{e^{\gamma'Z_{i}}}{1+e^{\gamma'Z_{i}}}.
\end{equation}
In this model (\ref{m2}), $\beta=(\beta_1, \ldots, \beta_p)'\in \mathbb R^p$ is an unknown regression parameter of interest ($\beta$ measures the association between the potential predictors $\mathbf X_i$ and the occurrence of the event of interest for a susceptible individual), and $\gamma=(\gamma_1, \ldots, \gamma_q)'\in \mathbb R^q$ is an unknown nuisance parameter. Letting $\theta:=(\beta',\gamma')'$, the log-likelihood for $\theta$ from the sample $\mathcal O_1,\ldots,\mathcal O_n$ (where $\mathcal{O}_i=(Y_i, \mathbf X_i, \mathbf Z_i)$, $i=1,\ldots,n$) is
	\begin{eqnarray}	\label{key3}
			l_{n}(\theta) &=&\sum_{i=1}^{n}\left[  J_{i}\log\left(e^{\gamma'Z_{i}}+(1+e^{\beta'X_{i}})^{-1}\right)-\log\left(1+e^{\gamma'Z_{i}}\right) 
			+ (1-J_{i})\left(Y_{i}\beta'X_{i}-log(1+e^{\beta'X_{i}})\right)\right]  \nonumber\\
			&=& \sum_{i=1}^{n}l_{i}(\theta)
		\end{eqnarray}	
	où $J_{i}:=\mathbf{1}_{\{Y_{i}=0\}}$.
\citep{ddd11} established the asymptotic properties of the maximum likelihood estimators of the parameters of a parametric ZIBer regression model.
\subsection{Regularization methods}
Regularization Techniques is an unavoidable and important step to improve the model prediction,  reduce errors and variable selection. This is also called the Shrinkage method. Which we use to add the penalty term to control the complex model to avoid overfitting by reducing the variance.
Choosing an appropriate value for the regularization parameter $\lambda$ is crucial in regularization regression, as it directly influences the bias-variance tradeoff and the overall performance of the model. Several methods (such as Cross-Validation, Generalized Cross-Validation, Information Criteria...) have been proposed for selecting the optimal ridge parameter, each with its own advantages and limitations. \\
We use the penalization function $\mathcal P(\theta,\lambda)$ defined by
\begin{equation}
		\mathcal{P}(\theta,\lambda)=n\left(\sum_{i=1}^{p}p(\lambda_{\beta},|\beta_{i}|)+\sum_{j=1}^{q}p(\lambda_{\gamma},|\gamma_{i}|)\right)
	\end{equation}
where $0<\lambda_{\beta}<1$ and $0<\lambda_{\gamma}<1$ represent the regularization parameters and control penalty intensity.\\
We use the following three penalty functions:
\begin{itemize}
		\item[$\blacktriangleright$]  For LASSO regression: $\mathcal{P}(\lambda,|\zeta|)=\lambda|\zeta|$, $\lambda \geq0$. 
		\item[$\blacktriangleright$] For ridge regression: $\mathcal{P}(\lambda,|\zeta|)=\lambda\zeta^{2}$, $\lambda \geq0$
		\item[$\blacktriangleright$] Foor Elastic-net regression: $\mathcal{P}(\lambda,|\zeta|)=\alpha \lambda|\zeta|  +\left(\frac{1-\alpha}{2}\right)\lambda\zeta^{2}$, $\lambda \geq0$.
	\end{itemize}
The penalized log-likelihood function is defined by
	\begin{equation}
	lp_{n}(\theta)	=l_{\text{pénalisé}}(\theta)=l_{n}(\theta)+\mathcal{P}(\theta,\lambda).
		\label{foncpen}
\end{equation}
The MLE $\widehat \theta_n:=(\widehat \beta_n',\widehat \gamma_n')'$ of $\theta$ is defined as the solution of the score equation 
\begin{equation}\label{eqs}
\dot{l}p_{n}(\theta) = \partial lp_n(\theta) \slash \partial \theta=0, 
\end{equation}
which can be solved, for example, using the \texttt{optim} function of the software \texttt R.
\subsubsection{LASSO regularization ($L_1$ Regularization)}
LASSO regression, also known as $L_1$ regularization is a regularization technique that applies a penalty to prevent overfitting and enhance the accuracy of statistical models (\cite{tibs96}). It is frequently used in machine learning to handle high dimensional data as it facilitates automatic feature selection with its application. It does this by adding a penalty term to the log-likelihood fonction, which is then multiplied by the regularization parameter ($\lambda$). This regularization parameter controls the amount of regularization applied. Larger values of lambda increase the penalty, shrinking more of the coefficients towards zero; this subsequently reduces the importance of (or altogether eliminates) some of the features from the model, resulting in automatic feature selection. Conversely, smaller values of lambda reduce the effect of the penalty, retaining more features within the model.\\
The log-likelihood function penalized by the $L_1$ norm is defined by
\begin{equation}\label{flasso}
lp_{n}^{\text{lasso}}(\beta,\gamma) = l_{n}(\beta,\gamma)+\lambda_{\beta}\sum_{i=1}^{p}|\beta_{i}|+\lambda_{\gamma}\sum_{i=1}^{q}|\gamma_{i}|.
\end{equation}
\subsubsection{Ridge regularization ($L_2$ Regularization)}
Ridge regression, also known as $L_2$ regularization, is a technique used in regression to address the problem of multicollinearity among predictor variables (\cite{ces92}). Multicollinearity occurs when independent variables in a regression model are highly correlated, which can lead to unreliable and unstable estimates of regression coefficients.\\
Ridge regression mitigates this issue by adding a regularization term to the log-likelihood objective function, which penalizes large coefficients and thus reduces their variance. This penalty discourages the model from using large values for the coefficients (the numbers multiplying the features). It forces the model to keep these coefficients small. 
The log-likelihood function penalized by the $L_1$ norm is defined by
\begin{equation}\label{fridge}
lp_{n}^{\text{ridge}}(\beta,\gamma) = l_{n}(\theta)+\lambda_{\beta}\sum_{i=1}^{p}\beta_{i}^{2}+\lambda_{\gamma}\sum_{i=1}^{q}\gamma_{i}^{2}
\end{equation}
\subsubsection{Elastic-net regularization}
The elastic net is a regularized regression method that has been widely used in learning and variable selection. The elastic-net regularization linearly combines an $L_1$ penalty term (like the lasso) and an $L_2$ penalty term (like ridge regression). The $L_1$ penalty term enforces sparsity of the elastic net estimator, whereas the $L_2$ penalty term ensures democracy among groups of correlated variables.
The log-likelihood function penalized is defined by
\begin{equation}\label{fenet}
lp_{n}^{\text{e-net}}(\beta,\gamma) =l_{n}(\beta,\gamma)+\lambda_{\beta}\alpha\left(\sum_{i=1}^{p}|\beta_{i}|+\sum_{i=1}^{q}|\gamma_{i}|\right)+\lambda_{\gamma}\frac{1-\alpha}{2}\left(\sum_{i=1}^{p}\beta_{i}^{2}+\sum_{i=1}^{q}\gamma_{i}^{2}\right)
\end{equation}
\begin{remark}
The regulation parameters $\lambda_\beta$ and $\lambda_\gamma$ can be determined on the basis of the $BIC$ (see \cite{schw78} for more details) defined by 
\begin{equation*}
BIC=-2lp_{n}(\theta)+log(n)\left[\sum_{i=1}^{p} 1_{\{ \beta_{i} \neq 0\}}+\sum_{j=1}^{q} 1_{\{\gamma_{i} \neq 0\}}+ \right]. 
\end{equation*}
\end{remark}

\section{Asymptotic theory}
\label{sec:asymptotic}
\subsection{Some further notations}
Define first the $(p\times n)$ and $(q\times n)$ matrices
\begin{equation*}
\mathbb X=
\begin{pmatrix}
1& 1 & \dots & 1\\
X_{12} & X_{22} &\dots & X_{n2}\\
\vdots & \vdots & \ddots & \vdots \\
X_{1p} & X_{2p} &\ldots & X_{np}
 \end{pmatrix}
 \quad
  \text{and} 
  \quad  
 \mathbb Z =
 \begin{pmatrix}
1& 1 & \dots & 1\\
Z_{12} & Z_{22} &\dots & Z_{n2}\\
\vdots & \vdots & \ddots & \vdots \\
Z_{1q} & Z_{2q} &\ldots & Z_{nq}
 \end{pmatrix}
\end{equation*}
and let $\mathbb W$ be the $((p+q)\times 2n)$ block-matrix defined as
\begin{equation*}
\mathbb W =
\begin{bmatrix}
\mathbb X & 0_{pn}\\
0_{qn} & \mathbb Z \\
\end{bmatrix}
\end{equation*}
where $0_{ab}$ denotes the $(a\times b)$ matrix whose components are all equal to zero (for any positive integer values $a$, $b$).~\\
Let $S(\theta)=(S_{j}(\theta))_{1\leq j \leq 2n}$ be the $2nn$-dimensional column vector defined as
\begin{equation*}
S(\theta)=\left(\dot{lp}_{\beta_{1}}(\theta),\dot{lp}_{\beta_{2}}(\theta), \dots,\dot{lp}_{\beta_{p}}(\theta) ,\dot{lp}_{\gamma_{1}}(\theta),\dot{lp}_{\gamma_{2}}(\theta)\dots\dot{lp}_{\gamma_{q}}(\theta)\right)',
\end{equation*}
where, for $i=1, \ldots, n$
\begin{align*}
 \dot{lp}_{\beta_{i}}(\theta)&=-J_{i}\frac{e^{\beta'X_{i}}Q_{i}(\theta)}{(1+e^{\beta'X_{i}})g_{i}(\theta)k_{i}(\theta)}+Z_{i}\left(\frac{e^{\beta'X_{i}}}{f_{i}(\theta)}+1\right)-\frac{e^{\beta'X_{i}}}{f_{i}(\theta)} 
+\frac{e^{\beta'X_{i}}Q_{i}(\theta)}{(1+e^{\beta'X_{i}})g_{i}(\theta)}+
\dot{\mathcal{P}}(\theta,\lambda)\\ 
 \dot{l}p_{\gamma_{i}}(\theta)&=  -J_{i}\frac{e^{\beta'X_{i}}+e^{\gamma'Z_{i}}}{g_{i}(\theta)k_{i}(\theta)}+Z_{i}\left(\frac{e^{\gamma'Z_{i}}}{f_{i}(\theta)}-1\right)-\frac{e^{\gamma'Z_{i}}}{f_{i}(\theta)} 
 +\frac{e^{\beta'X_{i}}+e^{\gamma'Z_{i}}}{g_{i}(\theta)} + \dot{\mathcal{P}}(\theta,\lambda) \\
 f_{i}(\theta)&=1+e^{\beta'X_{i}}+e^{\gamma'Z_{i}}\\
 g_{i}(\theta)&=e^{\beta'X_{i}}+e^{\gamma'Z_{i}}\\
 h_{i}(\theta)&=\frac{e^{\gamma^{T}Z_{i}}}{(1+e^{\beta^{T}X_{i}})g_{i}(\theta)}\\
 k_{i}(\theta)&= 1+g_{i}(\theta)h_{i}(\theta)\\
Q_{i}(\theta)&=1+e^{\beta^{T}X_{i}}-f_{i}(\theta)-e^{\beta^{T}X_{i}},
\end{align*}
and
	\begin{itemize}
	\item[$\blacktriangleright$] for LASSO method:
	\begin{equation*}
		 \dot{\mathcal{P}}(\lambda,\theta)
		=\left(\lambda_{\beta}\frac{\beta_{1}}{|\beta_{1}|}+ \lambda_{\gamma}\frac{\gamma_{1}}{|\gamma_{1}|},\dots, \lambda_{\beta}\frac{\beta_{p}}{|\beta_{p}|}+ \lambda_{\gamma}\frac{\gamma_{q}}{|\gamma_{q}|}                                          \right)'
	\end{equation*} 
	\item[$\blacktriangleright$] for ridge method:
	\begin{equation*}
		 \dot{\mathcal{P}}(\lambda,\theta)=2\left(\lambda_{\beta}\beta_{1}+ \lambda_{\gamma}\gamma_{1},\dots, \lambda_{\beta}\beta_{p}+ \lambda_{\gamma}\gamma_{q}      \right)'
		\end{equation*}
	\item[$\blacktriangleright$] for elastic-net method:
	\begin{equation*}
		\dot{\mathcal{P}}(\lambda,\theta)=\left(
		\begin{array}{c}
		\alpha\lambda_{\beta}\frac{\beta_{1}}{|\beta_{1}|}+ \alpha\lambda_{\gamma}\frac{\gamma_{1}}{|\gamma_{1}|}+(1-\alpha)\lambda_{\beta}\beta_{1}+(1-\alpha)\lambda_{\gamma}\gamma_{1}\\
		\vdots\\
		\alpha\lambda_{\beta}\frac{\beta_{p}}{|\beta_{p}|}+ \alpha\lambda_{\gamma}\frac{\gamma_{q}}{|\gamma_{q}|}+(1-\alpha)\lambda_{\beta}\beta_{p}+(1-\alpha)\lambda_{\gamma}\gamma_{q}
		 \end{array}
		\right) 
	\end{equation*}
\end{itemize}
Then, simple algebra shows that the score equation can be rewritten as
\begin{equation*}
\dot{l}p_n(\theta) = WS(\theta) = \sum_{j=1}^{2n} \mathbb W_{\bullet j}S_j(\theta) = 0.
\end{equation*}
where $\mathbb W_{\bullet j}$ is the $j$-th column $(j = 1,\ldots, p+q)$ of the matriw $\mathbb{W}$, that is, $\mathbb{W}_{\bullet j} = (W_{1j},\ldots, W_{(p+q)j})'$.~\\
We shall further note $\ddot{l}p_n(\theta)$ the $((p+q) \times (p+q))$ matrix of second derivatives of $lp_n(\theta)$, that is $\ddot{l}p_n(\theta) = \frac{\partial^2 lp_n(\theta)}{\partial\theta \partial\theta'}$. Let $\mathbf D(\theta)=(\mathbf D_{ij}(\theta))_{1\leq i,j \leq 2n}$ be the $(2n \times 2n)$ block matrix defined as
\begin{equation*}
\begin{bmatrix}
	D_{1}(\theta) & D_{0}(\theta)\\
	D_{0}(\theta) & D_{2}(\theta)
\end{bmatrix}
\end{equation*}
where $\mathbf D_1(\theta), \mathbf D_2(\theta)$ and $\mathbf D_0(\theta)$ are $(n\times n)$ diagonal matrices, with $i$-th elements $(i= 1,\ldots, n)$ respectively given by
\begin{align*}
	D_{1,ii}&=\frac{J_{i}e^{\beta^{T}X_{i}}}{k_{i}^{2}(\theta)}
	\left(k_{i}(\theta)-e^{\beta^{T}X_{i}}\left[2e^{\gamma^{T}Z_{i}}h_{i}(\theta)+1\right]\right)+\frac{(1-J_{i})e^{\beta X_{i}}}{h_{i}^{2}(\beta)}\\
	 D_{2,ii}&=\frac{J_{i}e^{\gamma^{T} Z_{i}}h_{i}^{2}(\beta)}{k_{i}^{2}(\theta)}\left(e^{\gamma^{T}Z_{i}}h_{i}^{2}(\beta)-k_{i}(\theta)\right)+\frac{e^{\gamma^{T}Z_{i}}}{1+e^{\gamma^{T}Z_{i}}}\\
	D_{0,ii}&=\frac{J_{i}exp(\beta^{T}X_{i}+\gamma^{T}Z_{i})h_{i}^{2}(\beta)}{k_{i}^{2}(\theta)} 
	\end{align*}
Then, some algebra shows that $\ddot{l}_n$ can be expressed as
\begin{equation*}
\ddot{l}_n(\theta) = -\mathbb{W}\mathbb{D}(\theta)\mathbb{W}'.
\end{equation*}
In the next section we establish the existency, consistency and asymptotic normality of the MLE of (\ref{foncpen}).
\section{Asymptotic results}
We first state some regularity conditions that will be needed to ensure identifiability.
\begin{description}
\item[C1] The covariates are bounded that is, there exist compact sets $F\subset\mathbb R^p$ and $G\subset\mathbb R^q$ such that $\mathbf X_i\in F$ and $\mathbf Z_i\in G$ for every $i=1,2,\ldots$ For every $i=1,2,\ldots$, $j=2,\ldots,p$, $k=2,\ldots,q$, $\mbox{var}[X_{ij}]>0$ and $\mbox{var}[Z_{ik}]>0$. For every $i=1,2,\ldots$, the $X_{ij}$
$(j=1,\ldots,p)$ are linearly independent, and the $Z_{ik}$ $(k=1,\ldots,q)$ are linearly independent.
\item[C2] Let $(\beta_{0}^{'}, \gamma_{0}^{'})^{'}$ denote the true parameter value. $\beta_0$ and $\gamma_0$ lie in the interior of known compact sets $\mathcal B \subset\mathbb R^p$ and $\mathcal G\subset\mathbb R^q$ respectively.
\item[C3] The Hessian matrix $\ddot{l}_n(\theta)$ is negative definite and of full rank, for every $n=1,2,\ldots$ Let $\lambda_n$ and $\Lambda_n$ be respectively the smallest and largest eigenvalues of $\mathbb W \mathbb D(\theta_0) \mathbb W'$. There exists a finite positive constant $c_2$ such that $\Lambda_n\slash\lambda_n<c_2$ for every $n=1,2,\ldots$
\item[C4] Let $A$ denote a symetric $(p+q)\times (p+q)$ matrix such that $\|A\| = \max_{1\leq i\leq k}|\lambda_i|$, where $(\lambda_i)_{1\leq i \leq k}$ are eigenvalues of matrix $A$.
\end{description}
The conditions \textbf{C1}, \textbf{C2} and \textbf{C3} are classical conditions for identifiability and asymptotic results in standard regression models (see, for example, \cite{gm81} and \cite{g01}). Note \cite{ddd11} and \cite{lo22} use a similary condition to prove the identifiability of the mixture model with logit link function for both models and the generalized extreme value regression model.
\begin{theorem}[Existence and consistency]\label{th1}
Under the conditions \textbf{C1}-\textbf{C5}, The penalized maximum likelihood estimator $\widehat \theta_n$ exists almost surely as $n\rightarrow \infty $ and converges almost surely to $\theta_0$, if and only if $\lambda_n$ tends to infinity as $n\rightarrow \infty$.
\end{theorem}

\section{A simulation study}
\label{sec:simul}
\subsection{Study design}
In this section, we investigate the numerical properties of the maximum likelihood estimator $\hat{\beta}_n$, under various conditions. We compare via simulations, the performance of the three regularizations regressions (LASSO, ridge and elastic-net). The simulation setting is as follows. We consider respectively the following models for the event of interest and the zero inflation part :
	\begin{align}\label{eqsim}
		logit(p_{i})&=\beta'X_{i}=\beta_{1}X_{i1}+\beta_{2}X_{i2}+\beta_{3}X_{i3}+\beta_{4}X_{i4}+\beta_{5}X_{i5} \nonumber\\
		logit(\pi_{i})&=\gamma'Z_{i}=\gamma_{1}Z_{i1}+\gamma_{2}Z_{i2}+\gamma_{3}Z_{i3}+\gamma_{4}Z_{i4}+\gamma_{5}Z_{i5}
	\end{align}
where $X_{i1}=Z_{i1}=1$ for each individual $i$ $(i=1,...,n)$. The covariates $X_{i2}$ and $Z_{i2}$ are independently drawn from normal $\mathcal{N}(0,1)$, normal $\mathcal{N}(-1,1)$ respectively. The covariates $X_{i3}$ and $Z_{i3}$ are independently drawn from Bernoulli $\mathcal{B}(1,0.9)$, normal $\mathcal{B}(1,0.5)$ respectively.  The covariates $X_{i4}$ and $Z_{i4}$ are independently drawn from uniform $\mathcal{U}[2,5]$, exponential $\mathcal{E}(1)$ respectively. The covariates $X_{i5}$ and $Z_{i5}$ are independently drawn from normal $\mathcal{B}(5,0.5)$, normal $\mathcal{E}(3)$ respectively. \\
We carry out our studies under two scenarios based on the percentages of zero inflation in the sample as follows:
\begin{description}
\item[$\bullet$]\textbf{Scenario 1}: The regression parameter $\theta=(\beta,\gamma)'$ is chosen as $\beta=(-0.9,-0.65,-0.2, 0.65, 0)'$ and $\gamma =(-0.55,-0.7,-1,0.45, 0)'$. In this setting, the average proportion of immunes data is $25\%$.
\item[$\bullet$]\textbf{Scenario 2}: The regression parameter $\theta=(\beta,\gamma)'$ is chosen as $\beta = (-0.9,-0.65,-0.2, 0.65, 0)$ and $\gamma = (0.25,-0.4, 0.8, 0.45, 0)'$. In this setting, the average proportion of immunes data is $50\%$.
\end{description}
 An i.i.d. sample of size $n\geq 1$ of the vector $(Y,X,Z)$ is generated from the model (\ref{eqsim}), and for each individual $i$, we get a realization $(y_i,x_i,z_i)$, . A maximum likelihood estimator $\hat{\theta}_n$ of $\theta = (\beta_{1},\beta_{2},\beta_{3}, \beta_{4}, \beta_{5},\theta_{1},\theta_{2},\theta_{3},\theta_{4}, \theta_{5})'$ is obtained from this dataset by solving the score equation (\ref{eqs}), using the \textbf{\texttt{optim}} function of the software \textbf{\texttt{R}}. The finite-sample behavior of the maximum likelihood estimator $\hat{\theta}_n$ was assessed for several sample sizes $(n = 100,500,1000)$ based on the two scenarios.
\subsection{Results}
For each configuration (sample size, percentage of zero inflation) of the design parameters, $N = 5000$ samples were obtained. Based on these $N=5000$ replicates, we obtain averaged values for the estimates of the parameters $\theta_j$, $j=1,\ldots,10$, which are calculated as $N^{-1}\sum_{k=1}^{N}\hat{\theta}_{j,n}^{(k)}$, where $\hat{\theta}_{j,n}^{(k)}$ is the estimate obtained from the $k$-th simulated sample. The quality of estimates is evaluated by using using the empirical coverage probability and average length of 95\%-level confidence interval (CI), standard deviation (SD) and error (SE), Bias and the Root Mean Square Error (RMSE) defined as, for $j=1, ..., 10$
\begin{equation*}
\begin{split}
\text{Bias}(\hat{\theta}_{n,j}) &= \mathbb{E}(\hat{\theta}_{n,j} -\theta)~ \approx ~\frac{1}{N}\sum_{k=1}^{N}\left( \hat{\theta}_{j,n}^{(k)} -\theta\right) \\
\text{RMSE}(\hat{\theta}_{n,j}) &= \sqrt{\mathbb{E}\left[ (\hat{\theta}_{n,j} -\theta)^2\right]}~ \approx ~\sqrt{\frac{1}{N}\sum_{k=1}^{N}\left( \hat{\theta}_{j,n}^{(k)} -\theta\right) ^2}
\end{split}
\end{equation*}
\begin{table}[h]
		\centering
		\begin{tabular}[h]{|l|c|c|c|c|c|}
			\hline
			\backslashbox{$\lambda$ }{Methods} & LASSO & Ridge& Elastic-net \\
			\hline \hline    
			0.0001 & 0.000874 & 0.000823 & 0.000751  \\ 
			0.0010 & 0.000793 & 0.000709 & 0.000625 \\ 
			0.0100&  0.000168 & 0.000155 & 0.000107  \\ 
			0.0500&  0.000009 & 0.000008 & 0.000000 \\ 
			0.0900&  0.000097 & 0.000087 & 0.000080  \\ 
			0.1000&  0.000005 & 0.000003 & 0.000002 \\ 
			0.3000&  0.000017 & 0.000011 & 0.000009 \\
			0.5000&  0.000059 & 0.000056 & 0.000041  \\
			10.000&  0.000755 & 0.000635 & 0.000513  \\
			1000.0&  0.000981 & 0.000810 & 0.000780 \\ 
			\hline
		\end{tabular}
		\caption{Choice of regularization parameter $\lambda$ for $n$ fixed at $1000$ by calculating $RMSE$. }
		\label{t1}
	\end{table} 
For each configuration \texttt{sample size} $\times$ \texttt{percentage of zero inflation} of the design parameters, $N=5000$ samples are obtained. Based on these $N$ repetitions, we obtain averaged values for the estimates of the  $\theta_l$ $(l=1,\ldots,10)$, which are calculated as $N^{-1}\sum_{j=1}^N \widehat \theta_{l,n}^{(j)}$, where $\widehat \theta_n^{(j)}=(\widehat \beta_{1,n}^{(j)}, \ldots, \widehat \beta_{5,n}^{(j)}, (\widehat \gamma_{1,n}^{(j)}, \ldots, \widehat \gamma_{5,n}^{(j)})'$ is the estimate obtained from the $j$-th simulated sample. For each of the parameters $\theta_l$, we also obtain the empirical coverage probability and average length of $95\%$-level confidence interval (CI), standard deviation (SD) and error (SE), Bias, empirical root mean square (RMSE) and mean absolute errors (MAE), based on the $N$ samples. \\
Tables \ref{t2}-\ref{t7} give the results for the two scenarios respectively. 
~\\~\\
From Tables \ref{t2} to \ref{t7}, for the different values of $n$ chosen, we can see that all three methods (LASSO, Ridge and Elastic-net) provide a reasonable approximation of the true regression parameter, even when the zero inflation fraction is high. Indeed, the bias, RMSE, SE and SD become progressively smaller as the sample size increases. Overall, the bias, RMSE, SE and SD of $\widehat \beta_n$ and $\widehat \gamma_n$ stay limited while its variability increases with the percentage of zero inflation. This increase is particularly noticeable when the sample size is small ($n=100$).
\par In almost all tables, the empirical coverage probabilities are close to the confidence level of $95\%$. Empirical coverage probabilities are also close to $1$, as the sample size increases. At the same time, for a fixed $n$, we observe that parameter performance remains stable. These observations illustrate the general fact that accurate estimation in a zero-inflation regression model requires a balance between susceptible and non-susceptible subpopulations (i.e. a sufficient quantity of zero and non-zero observations must be available to accurately estimate zero-inflation probability).
\par From the tables obtained, we can say that the estimates obtained by the Elastic-net method are better than those obtained by the Ridge method, which is still better than those obtained by the LASSO method. In fact, estimates of the ZIBer model using the LASSO, Ridge and Elastic-net regression methods have more or less similar performances. The RMSE, bias, SD and SE of the Elastic-net method are lower than those of the Ridge method, which are even lower than those of the LASSO method. The coverage probabilities are close to the nominal confidence level, indicating that the asymptotic variances are correctly estimated. The bias of the estimates of the different methods decreases considerably as the sample size increases, but the empirical coverage probabilities are close to the nominal confidence level, even if there is a misspecification of the probability of success. Empirical coverage probabilities get closer and closer to $1$ as sample size increases.
\par The results obtained in these tables confirm those in Table 8, where the log-likelihood function and AIC are calculated. In particular, the Elastic-net method remains better and more accurate than the LASSO and Ridge methods.\\
\par Through simulations, we also investigate the quality of the normal approximation of the asymptotic distribution of the interest parameter estimator $\hat{\beta}_{n,j}$ in the model (\ref{eqsim}). For each configuration ($n\in \{100, 500, 1000\}$ and percentage of zero inflation in $n\in \{25\%, 50\%\}$) of the design parameters, we obtain the histograms of the $\hat{\beta}_{n,j}^{(k)}, k=1,..., N$ and the corresponding QQ-plots. The graphs are displayed in Figures \ref{fig1} to \ref{fig6}. From these figures, the normal approximation, as expected is reasonably even if the sample size and the proportion of zero-inflation seem low. These findings are coherent with our previous observations for $\hat{\beta}_{j,n}$ especially when the sample size is small ($n$ around $100$, say). 
%
\begin{table}
		\centering
		\rotatebox{90}{
			\begin{tabular}{|l|c|c|c|c|c|c|c|c|c|c|c|}
				\hline
				Estimater  &\multicolumn{10}{c|}{LASSO}\\ \cline{2-11}  
				&\multicolumn{5}{c|}{Logistic component}&\multicolumn{5}{c|}{Zero component}\\ 
				\cline{2-11}  
				& $\hat{\beta_{1n}}$  &  $\hat{\beta_{2n}}$    & $\hat{\beta_{3n}}$  &  $\hat{\beta_{4n}}$ &  $\hat{\beta_{5n}}$ 	& $\hat{\gamma_{1n}}$  &  $\hat{\gamma_{2n}}$    & $\hat{\gamma_{3n}}$  &  $\hat{\gamma_{4n}}$ &  $\hat{\gamma_{5n}}$  \\\hline
				
				bias    & -0.001008 & -0.000967&  0.000975 & 0.001152 & 0.008919& -0.009387 & -0.008789  &0.007998& -0.009887& -0.007896\\
				rel.bias& -0.100814 & -0.096743&  0.097473 & 0.111526 & 0.089913& -0.938722 & -0.878918  &0.799827& -0.988654& -0.789611  \\
				SD      &  0.011110 &  0.025370&  0.016060 & 0.016970 & 0.009820&  0.098879 &  0.019820  &0.776760&  0.098850&  0.090350
				                     \\ 
				SE      &  0.001111 &  0.002537&  0.001606 & 0.001697 & 0.000982&  0.0098879&  0.001982  & 0.077676&  0.009885 &  0.009035      \\
				RMSE    &  0.000875 &  0.000909&  0.000975 & 0.000968 & 0.001958&  0.0019368&  0.032203  & 0.065440&  0.050175 &  0.031684 \\
             	l(CI)   &6.43e-05   &  < 2e-16 & < 2e-16   & < 2e-16  & 1.41e-13&  0.171125 &  0.779802  & 0.451896 & 0.677465 &  0.747564  \\
             	CP      &  0.986344 &  0.984705&  0.94540  & 0.986881 & 0.983961&  0.9802368&  0.980132  & 0.980505 & 0.988108 &  0.989332  \\         

				\hline
				\hline
				&\multicolumn{10}{c|}{Ridge}\\ \cline{2-11}
				bias    & 0.000166 &-0.000176 & -0.000131  & 0.000165&  0.000187& -0.000176& -0.000175   & 0.000188 & -0.000197 &-0.000199     \\
				rel.bias& 0.016623 &-0.017626 & -0.013104  & 0.016525&  0.018722& -0.017624& -0.017546   & 0.018803 & -0.019766 &-0.019952  \\
				SD      & 0.008871 & 0.008257 &  0.007572  & 0.008031&  0.008191&  0.009941&  0.009815   & 0.009434 &  0.009142 & 0.009153 \\
				SE      & 0.000887 & 0.000825 &  0.000757  & 0.000803&  0.000819&  0.000994&  0.000981   & 0.000943	&  0.000914 & 0.000915 \\
				RMSE    & 0.000098 & 0.000156 &  0.000182  & 0.000135&  0.000146&  0.000199&  0.000272   & 0.000287	&  0.000237 & 0.000211 \\
				  l(CI) &< 2e-16 &0.000169    &  7.05e-12  & NaN     &  2.34e-09&  0.585457&  0.504791   & 0.144783 &  0.656151 & 0.135220  \\ 
				CP      & 0.987377  &0.988389 &  0.985701  & NaN     &  0.987873&  0.980546&  0.971664   & 0.975144 &  0.961953 & 0.959254  \\
				
				\hline
				\hline
				&\multicolumn{10}{c|}{Elastic net}\\ \cline{2-11}
				bias    &-0.000114 & 0.000103  &  0.000108 &  0.000145& 0.000153 &  0.000171 &-0.000173 & 0.000179  &-0.000193 & -0.000109   \\
				rel.bias&-0.011439 & 0.010335  &  0.010448 &  0.014555& 0.015396 &  0.017077 &-0.017334 & 0.017923  &-0.019319 & -0.010941  \\         
				SD      & 0.007421 & 0.007424  &  0.005821 &    NaN   & 0.008526 &  0.008572 & 0.009651 & 0.008668  & 0.008631 &  0.007690  \\
				SE      & 0.000742  & 0.000742 & 0.000582 & NaN      &0.000852  &0.000857  & 0.000965 &0.000866    &0.000861  &0.000769  \\
				RMSE    & 0.000038 &  0.000039&  0.000029B& 0.000027 &0.000024 & 0.000043  & 0.000055 &0.000063   & 0.000067  &0.000069 \\
				l(CI)   &< 2e-16  & 0.001169 &7.05e-12 &  NaN & 2.34e-09& 0.585004  &0.501290  &0.198483 & 0.196151   & 0.149221  \\
				CP      &0.998273 & 0.989989 &  0.998985 &NaN  &0.996973 &0.979467 &0.989660 & 0.987446 & 0.978503 & 0.970547   \\
				\hline
			\end{tabular}
		}
		\caption{Simulation results for scenario 1 ($25\%$ zero-inflation proportion) with $n = 100$}
		\label{t2}
	\end{table}
	\begin{table}
		\centering
		\rotatebox{90}{
			\begin{tabular}{|l|c|c|c|c|c|c|c|c|c|c|c|}
				\hline
				Estimater  &\multicolumn{10}{c|}{LASSO}\\ \cline{2-11}  
				&\multicolumn{5}{c|}{Logistic component}&\multicolumn{5}{c|}{Zero component}\\ 
				\cline{2-11}  
				& $\hat{\beta_{1n}}$  &  $\hat{\beta_{2n}}$    & $\hat{\beta_{3n}}$  &  $\hat{\beta_{4n}}$ &  $\hat{\beta_{5n}}$ 	& $\hat{\gamma_{1n}}$  &  $\hat{\gamma_{2n}}$    & $\hat{\gamma_{3n}}$  &  $\hat{\gamma_{4n}}$ &  $\hat{\gamma_{5n}}$  \\\hline
				bias    & 0.000037 & -0.000038 &-0.000035& 0.000014 & 0.000039& -0.000045	& -0.000041 & 0.000048& -0.000047 &-0.000051    \\
				rel.bias& 0.003712 & -0.003797 &-0.003564& 0.001433 & 0.000976& -0.000515  & -0.000343 & 0.000824& -0.000679 &-0.000104    \\
				SD      & 0.007244 &  0.001811 & 0.009704& 0.012946 & 0.014021&  0.019319  &  0.011336 & 0.017932 & 0.018223 & 0.01791\\
				SE      & 0.000324  & 0.000081 & 0.000434 & 0.000579 & 0.000627& 0.000864  &  0.000507 & 0.000802 & 0.000815 & 0.000801 \\
				RMSE    & 0.000030  & 0.000026 & 0.000016 & 0.000029 & 0.000022& 0.000033  &  0.000391 & 0.000405  &0.000112 & 0.000103 \\
				l(CI)   & 0.173407  & 0.21367  & 0.069259 & 0.593509 & 0.00735567& 0.00828623  &  0.0057052 & 0.001337  &0.001663 & 0.002189 \\
				CP      & 0.998405  & 0.9984478 & 0.988395 & 0.988534 & 0.999338& 0.977821  &  0.988108 & 0.973447  &0.977384 & 0.977691  \\
				
				\hline
				\hline
				&\multicolumn{10}{c|}{Ridge}\\ \cline{2-11}
				bias    & 0.000022 & -0.000024  & -0.000023 & 0.000015&  0.000022&-0.000017  &-0.000055&  0.000058 &-0.000067 &-0.000099     \\ 
				rel.bias& 0.002216 & -0.002441  & -0.002311 & 0.001451&  0.001220&-0.001764  &-0.005545&  0.005812 &-0.006714 &-0.009917 \\
				SD      & 0.007244 &  0.001811  &  0.009704 & 0.012946&  0.014012& 0.019319  & 0.011336&  0.017933 & 0.018223 & 0.017911 \\
				SE      &0.000241  &  0.000211  &  0.000255&  0.000242 & 0.000285& 0.000299  & 0.000382&  0.000342 & 0.000321 & 0.000685 \\
				RMSE    &0.000028  &   0.000027 &  0.000023&  0.000025 & 0.000018& 0.000037  & 0.000031&  0.000038 & 0.000037 & 0.000039 \\   
				l(CI)   &0.509158  &< 2e-16     &< 2e-16 &< 2e-16 &6.79e-14  &0.064542 &0.041005 &0.000204&0.256728& 0.840822 \\
				CP      &0.99886  1&   0.998380 &  0.997187&  0.988644 & 0.987492&0.988001  &  0.980438  &0.970714 & 0.972134 & 0.972201 \\
				\hline
				\hline
				&\multicolumn{10}{c|}{Elastic net}\\ \cline{2-11}
				bias    & 0.000019  & 0.000018  &  0.000013  &  0.000017 & 0.000020& -0.000032& -0.000038& 0.000074 & -0.000032 &-0.000061    \\
				rel.bias& 0.001930  & 0.001787  &  0.001291  &  0.001774 & 0.002077& -0.003235& -0.003799& 0.007424 & -0.003210 &-0.006107 \\
				SD      & 0.004964  & 0.005165  &  0.005567  &  0.004762 & 0.004561&  0.007065&  0.007133& 0.010084 &  0.008564 & 0.015921
				  \\
				SE      & 0.000222  & 0.000231  &  0.000249  &  0.000213 & 0.000204&  0.000316 & 0.000319 & 0.000451 & 0.000383  & 0.000712  \\
				RMSE    & 0.000017  & 0.000016  &  0.000013  &  0.000018 & 0.000021&  0.000018 & 0.000028 & 0.000026 & 0.000031  & 0.000033  \\
				l(CI)   &< 2e-16    &< 2e-16    &  4.99e-12  &< 2e-16    &< 2e-16 &0.520708    &0.000550  & 0.000528 & 0.202144  & 0.729068  \\
				CP      &0.999037   & 0.999240  &  0.998206  &  0.998103 & 0.999363 &  0.987642&0.975104  & 0.976603 & 0.970125  & 0.954346   \\
				\hline
			\end{tabular}
		}
		\caption{Simulation results for scenario 1 ($25\%$ zero-inflation proportion) with $n = 500$}
		\label{t3}
	\end{table}
	\begin{table}
		\centering
		\rotatebox{90}{
			\begin{tabular}{|l|c|c|c|c|c|c|c|c|c|c|}
				\hline
				Estimater  &\multicolumn{10}{c|}{LASSO}\\ \cline{2-11}  
				&\multicolumn{5}{c|}{Logistic component}&\multicolumn{5}{c|}{Zero component}\\ 
				\cline{2-11}  
				& $\hat{\beta_{1n}}$  &  $\hat{\beta_{2n}}$    & $\hat{\beta_{3n}}$  &  $\hat{\beta_{4n}}$ &  $\hat{\beta_{5n}}$ 	& $\hat{\gamma_{1n}}$  &  $\hat{\gamma_{2n}}$    & $\hat{\gamma_{3n}}$  &  $\hat{\gamma_{4n}}$ &  $\hat{\gamma_{5n}}$  \\\hline
				bias    &0.000011&-0.000012& -0.000015 & 0.000019& 0.000013& -0.000075& -0.000072& 0.000082& -0.000087& -0.000082 \\ 
				rel.bias&0.001146&-0.001098& -0.001527 & 0.001916& 0.001332& -0.007528& -0.007213& 0.008237& -0.008743& -0.008233 \\
				SD      & < 6.32e-15 &2.28e-03&          NaN& 6.08e-04 &<6.32e-15& 3.16e-05& 3.47e-04	& 6.64e-04 &7.27e-04 &1.38e-02  \\
				SE      &< 2e-16 & 7.22e-05 & NaN       &0.000019 &< 2e-16  & 0.000001&  0.000011& 0.000021&  0.000023&  0.000439 \\
				RMSE    &0.000005& 0.000004& 0.000009   & 0.000006& 0.000004& 0.000009&0.000082& 0.000015&  0.000093&  0.000082 \\
				l(CI)   &0.000836&-0.000239&  NaN       & 0.002934&-0.000121& 0.000539 &-0.000406&0.000038 & 0.000009&  0.000941  \\
				CP      &0.999873& 0.999364&  NaN       & 0.996098&0.9998049  &0.9898251 &0.989855 & 0.980239& 0.988307&  0.989086    \\

				\hline
				\hline
				&\multicolumn{10}{c|}{Ridge}\\ \cline{2-11}
				bias    &0.000001&-0.000008&-0.000007   & 0.000009 & 0.000004& -0.000016 & -0.000025 & 0.000028& -0.000027& -0.000029    \\
				rel.bias&0.000106&-0.000826&-0.000711   & 0.000916 & 0.000422& -0.001631 & -0.002553 & 0.000284& -0.000271& -0.000289\\
				SD      &0.000727& 0.000664& 0.000569   & 0.000316 & 0.000505&  0.001739 &  0.001834 & 0.001992&  0.002561&  0.002656\\
				SE      &0.000023& 0.000021& 0.000018  &  0.000010 & 0.000016&  0.000055 &  0.000058 & 0.000063&  0.000081&  0.000084\\
				RMSE    &0.000004& 0.000005& 0.000003  &  0.000008 & 0.000005&  0.000015 &  0.0000192& 0.000028&  0.000037&  0.000052  \\
			  	  l(CI) & 0.000014& 0.000273 &0.000857 &   0.000037 &0.000042 & 0.000369  & 0.0000895& 0.0000102 &0.000118&  0.000008\\
			  CP    	& 0.999898& 0.999906& 0.999214 &   0.999379 &0.999774 & 0.989902  & 0.9976141 & 0.981157 &0.987353 & 0.987137   \\
				\hline
				\hline
				&\multicolumn{10}{c|}{Elastic net }\\ \cline{2-11}
				bias     &0.000012&-0.000005 &-0.000004  &  0.000003 & 0.000001 & -0.000153	& -0.000018 & 0.000012 &-0.000024 &-0.000027   \\
				rel.bias &0.001243&-0.000483 &-0.000434  &  0.000329 & 0.000073 & -0.001536 & -0.001784 & 0.001236 &-0.002424 &-0.002697 \\
				SD       &0.000121& 0.000221 & 0.000126  &       NaN & 0.000221 &  0.001541 &  0.002687 & 0.002434 & 0.002941 & 0.003067   \\
				SE       &0.000004& 0.000007 & 0.000004  & NaN       & 0.000007 &  0.000049 &  0.000085 & 0.000077  &0.000093 & 0.000097  \\
				RMSE     &0.000001& 0.000003& 3e-16  &  0.000005 & 0.000006&  0.000012 &  0.0000032& 0.000015&  0.000010&  0.000013  \\
				CP       &< 2e-9 & 0.046292  &1.19e-08  &NaN   &6.37e-10  &0.117537 &7.90e-06 &0.000435 &0.001950 &0.367727   \\
				CP       &0.999927& 0.99993  & 0.999900  &NaN         &0.999998  &  0.99995  & 0.999914 & 0.999902   &0.999098  & 0.999901  \\
				\hline
				
			\end{tabular}
		}
		\caption{Simulation results for scenario 1 ($25\%$ zero-inflation proportion) with $n = 1000$}
		\label{t4}
	\end{table}

		\begin{table}
		\centering
		\rotatebox{90}{
			\begin{tabular}{|l|c|c|c|c|c|c|c|c|c|c|c|}
				\hline
				Estimater  &\multicolumn{10}{c|}{LASSO}\\ \cline{2-11}  
				&\multicolumn{5}{c|}{Logistic component}&\multicolumn{5}{c|}{Zero component}\\ 
				\cline{2-11}  
				& $\hat{\beta_{1n}}$  &  $\hat{\beta_{2n}}$    & $\hat{\beta_{3n}}$  &  $\hat{\beta_{4n}}$ &  $\hat{\beta_{5n}}$ 	& $\hat{\gamma_{1n}}$  &  $\hat{\gamma_{2n}}$    & $\hat{\gamma_{3n}}$  &  $\hat{\gamma_{4n}}$ &  $\hat{\gamma_{5n}}$  \\\hline
				bias    & 0.002853 & -0.006428 &  0.005478 & 0.004469 & 0.001625 & 0.001293 &-0.017013&-0.034301 & -0.009574 &  0.005438    \\ 
				rel.bias& 0.285355 & -0.642804 &  0.547863 & 0.446946 & 0.162506 & 0.129394 &-1.701306&-0.343035 & -0.957420 &  0.443841    \\
				SD      & 1.680231 & 0.52930   & 1.681424  &2.133453&  1.032703 & 0.080298 & 0.241370& 0.470312  & 8.755452 &  2.582097\\                                                      
				SE      & 0.168023 &  0.052930 &  0.168142 & 0.213345 & 0.103270 & 0.008029 & 0.024137& 0.047031 & 0.875545 & 0.258209   \\
				RMSE    & 0.331780 &  0.903932 &  0.327303 & 0.856551 & 0.654383 & 1.296589 & 0.362413& 0.583089 & 0.269333 & 0.071334  \\
				l(CI)   & 0.753449 &  0.435637 &  0.203024  &0.389500  &0.143901 & 0.632789 & 0.518097& 0.538789 & 0.616459 & 0.897001   \\
				 CP     & 0.893350 &   0.793971&   0.539167 & 0.904533&  0.850010 & 0.902743 & 0.811898&0.811154& 0.906045& 0.769990   \\
				\hline
				\hline  
				&	\multicolumn{10}{c|}{Ridge}\\ \cline{2-11}
				bias    & 0.000485  &  0.001075 & -0.002009 & 0.000367  & 0.001815 &-0.001319 &-0.001056&-0.001006&-0.001595  &  0.001365    \\
				rel.bias&0.0485428  &  0.107465 & -0.200944 & 0.036719  & 0.181509 &-0.131914 &-0.105640&-0.100633&-0.159541  &  0.136531  \\
				SD      &1.0677801  &  2.164110 & 1.548760  & 2.588370  & 3.445730 & 2.088192 & 0.576530& 0.849690& 1.201750  &  1.190950 \\
				SE      &0.106778   &  0.216411 & 0.154876  & 0.258837  & 0.344573 & 0.2088192& 0.057653& 0.084969& 0.120175  &  0.119095 \\
				RMSE    &0.098864   &  0.058656 & 0.064845  & 0.196034  & 0.102606 & 0.041572 & 0.154562& 0.089969&	0.109739  &  0.151917  \\
				l(CI)   &-0.82223   & -1.711435 &-0.306568  & 0.221568  &-0.767028 & 0.482681 &-0.097009&-0.115508&-0.968125  & -0.448234   \\
				CP      &0.794109   &  0.687028 & 0.759663  & 0.526332  & 0.666515 & 0.838345 & 0.789601& 0.875184& 0.890491  &  0.781476  \\

				\hline
				\hline
				&	\multicolumn{10}{c|}{Elastic net }\\ \cline{2-11}
				bias    &0.001086   &  0.001275 & 0.001064 & 0.001040   & 0.001469 & 0.001433 &-0.000993& -0.000985&-0.001004 & 0.000997    \\
				rel.bias&0.108568   &  0.127549 & 0.106440 & 0.104039   & 0.146911 & 0.143285 &-0.099485& -0.098536&-0.014043 & 0.099748 \\
				SD      &NaN        &  1.141233 &   NaN    & 1.247904   &     NaN  & 0.906414 & 0.955328&  1.067792& 0.813334 & 0.963241  \\
				SE      & NaN       & 0.114123  &      NaN & 0.124790   &       NaN& 0.090646& 0.095532 & 0.106779 & 0.081333 & 0.096324 \\
				RMSE    & 0.091492 &  0.083082  & 0.189959 & 0.099967   & 0.102306 & 0.109515& 0.139511 & 0.121741 & 0.098323 & 0.089017 \\
				l(CI)   &-0.345627 & -0.235996  & 0.108209 & 0.208207   &-0.550345 & 0.216919& NaN      &-0.114569 &-0.345954 &-0.127561     \\
				CP      & 0.886531 &  0.904645 &  0.914350 & 0.897227   & 0.900858 & 0.987972& NaN      & 0.897091 & 0.997412 & 0.987016   \\
				\hline
			\end{tabular}
		}
		\caption{Simulation results for scenario 2 ($50\%$ zero-inflation proportion) with $n = 100$.}
		\label{t5}
	\end{table}
		\begin{table}
		\centering
		\rotatebox{90}{
			\begin{tabular}{|l|c|c|c|c|c|c|c|c|c|c|c|}
				\hline
				Estimater  &\multicolumn{10}{c|}{LASSO}\\ \cline{2-11}  
				&\multicolumn{5}{c|}{Logistic component}&\multicolumn{5}{c|}{Zero component}\\ 
				\cline{2-11}  
				& $\hat{\beta_{1n}}$  &  $\hat{\beta_{2n}}$    & $\hat{\beta_{3n}}$  &  $\hat{\beta_{4n}}$ &  $\hat{\beta_{5n}}$ 	& $\hat{\gamma_{1n}}$  &  $\hat{\gamma_{2n}}$    & $\hat{\gamma_{3n}}$  &  $\hat{\gamma_{4n}}$ &  $\hat{\gamma_{5n}}$  \\\hline
				bias    &0.001972& 0.001007 &-0.001324&  0.001449 &  0.002909& -0.021900&-0.008920& -0.005985& -0.010462& -0.008546    \\
				rel.bias&0.197241& 0.100661& -0.132385&  0.144935 &  0.290924& -0.219009&-0.189201& -0.598527& -1.046190& -0.854630 \\ 
				SD      &2.453727& 0.626121&  2.475037&  2.195975 &  2.422131&  4.583000& 3.988117&  2.760270&  1.676470&  2.382933  \\
				SE      &0.109734 &0.028001 & 0.110687&  0.098207 &  0.108321&  0.204958& 0.178354& 0.123443&  0.074974&  0.106568 \\
				RMSE    &0.212739 &0.257075&  0.362192&  0.270662 &  0.140463 & 0.168565& 0.293867& 0.082534  &0.086290&  0.061404 \\
				l(CI)   &0.315001 &-0.038045 & 0.159823  &0.927370  &-0.819185 &-0.273001 &-4.233  &-0.011698 & 0.038322 &-0.775987  \\
				CP      &0.852862 &0.950725 & 0.949927 & 0.954039  & 0.813005 & 0.984892 &0.901092& 0.886072 & 0.901966  &0.938231 \\
				\hline
				\hline
				&\multicolumn{10}{c|}{Ridge}\\ \cline{2-11}
				bias    &0.001035 & -0.001039& -0.001227&  0.002134 &  0.002411 &-0.002101&-0.001853&-0.000938&-0.000993&-0.000994     \\
				rel.bias&0.103528 & -0.103894& -0.122717&  0.213425 & 0.241089&   0.210104& -0.18539&-0.093798&-0.992846&-0.994157\\
				SD      &2.275087 &  0.284495&  0.238767&  0.426799 & 0.247645&   2.080304&  2.19318& 2.216368& 2.200134& 1.993499   \\
				SE      &0.101745 & 0.012723 & 0.010678  & 0.019087 & 0.011075  & 0.093034  &0.098082 &0.099119  & 0.098393 &0.089152   \\
				RMSE    &0.100145 & 0.127845 & 0.129883 &  0.127831 & 0.100822  & 0.170170 & 0.126806&0.011468& 0.019981 & 0.009687  \\
				l(CI)      &< 2e-16  & 2.37e-13& 0.076420&0.000345 &< 2e-16 &0.035309 &5.01e-06 &0.000527 & 0.069732&0.215727   \\
				CP      &0.960901 & 0.961326 & 0.977044 &  0.981279 &  0.899567 & 0.970105 & 0.970565 &0.921467  &  0.899389 & 0.986238   \\    
				\hline
				\hline
				&\multicolumn{10}{c|}{Elastic net}\\ \cline{2-11}
				bias    &-0.000931 & 0.001004&0.001129&  0.001293 &  0.001241&-0.002004& -0.001807& -0.000918& -0.001023& -0.000906     \\
				rel.bias&-0.093098 & 0.100415&0.112939&  0.129331 &  0.124147&-0.200447& -0.180725&-0.0911791& -0.102366&-0.090630\\
				SD      & 2.451513 & 0.047628&0.148229&  0.270206 &  2.471101& 2.498045&  2.053247& 0.0374988&  0.353031& 0.043044   \\
				SE      &0.109635 &0.002130 &0.006629 &0.012084 &  0.110511&0.111716 & 0.091824 &0.001677 & 0.015788 & 0.001925   \\
				RMSE    & 0.090633 & 0.089624&0.109452&  0.102077 &  0.089687& 0.114781& 0.089301& 0.122816	& 0.005727& 0.019954  \\
				CP& 0.780929 & 0.000157&0.152159  &0.004318 &0.158585 &0.152206  &0.001965 &0.003394 &0.647585 & 0.196129  \\
				CP      &0.990542 & 0.977780 &0.899943 &0.985014 & 0.99410 &0.89432   &0.894095 &0.900039 &0.890457    &   0.887293 \\
				
				\hline
			\end{tabular}
	}
		\caption{Simulation results for scenario 2 ($50\%$ zero-inflation proportion) with $n = 500$}
		\label{t6}	
	\end{table}
	\begin{table}
		\centering
		\rotatebox{90}{
			\begin{tabular}{|l|c|c|c|c|c|c|c|c|c|c|}
				\hline
				Estimator  &\multicolumn{10}{c|}{LASSO}\\ \cline{2-11}  
				&\multicolumn{5}{c|}{Logistic component}&\multicolumn{5}{c|}{Zero component}\\ 
				\cline{2-11}  
				& $\hat{\beta_{1n}}$  &  $\hat{\beta_{2n}}$    & $\hat{\beta_{3n}}$  &  $\hat{\beta_{4n}}$ &  $\hat{\beta_{5n}}$ 	& $\hat{\gamma_{1n}}$  &  $\hat{\gamma_{2n}}$    & $\hat{\gamma_{3n}}$  &  $\hat{\gamma_{4n}}$ &  $\hat{\gamma_{5n}}$  \\\hline
				bias    & 0.001516&-0.001033&-0.001838& 0.001251& 0.001369 & -0.001805& -0.001569&-0.001198& -0.001687& -0.003881    \\ 
				rel.bias& 0.101648&-0.103326&-0.183832& 0.125148& 0.136904 & -0.180483& -0.156903&-0.119777& -0.168702& -0.388093  \\
				SD      & 1.664971& 8.767826& 4.579863& 8.923030& 4.595327 &  1.225351&  3.855322& 6.855944&  3.791950& 3.217333 \\
				SE      &0.052651 & 0.277263&0.144828 &0.282171 & 0.145317 &  0.038749 & 0.121916& 0.216804&  0.119912& 0.101741  \\
				RMSE    &0.003547 & 0.068072&0.166551 &0.088492 & 0.182073 &  0.019924 & 0.139758& 0.085335&  0.062297& 0.060926  \\
				l(CI)   &-0.3864  & 0.394233&-0.441266&0.496005 &-0.318097 &  0.425569 &-0.392563&-0.370159 & 0.409573& 0.369058  \\
				CP      &0.972712 & 0.947017&0.974385 &0.971257 & 0.975056 &  0.915428 & 0.977891& 0.952401 & 0.933054& 0.929001\\
				
				\hline
				\hline
				&\multicolumn{10}{c|}{Ridge}\\ \cline{2-11}
				bias    &0.001303 &-0.001365& -0.001526&0.001306& 0.001225 & -0.010209 &-0.001818&-0.001161&-0.001587& -0.000927   \\ 
				rel.bias&0.100359 &-0.136554& -0.152589&0.130607& 0.122531 & -0.102092 &-0.181802&-0.111121&-0.158721& -0.092701 \\
				SD      &0.521680 & 0.530819&  0.568767&0.389055& 0.617814 &  0.712619 & 2.704759& 2.706719& 2.504365&  0.687257\\
				SE      &0.016497 & 0.016786&  0.017986&0.012303& 0.019537&    0.022535& 0.085532& 0.085594& 0.079195 &  0.021733 \\
				RMSE    &0.002759 & 0.019525& 0.146666  & 0.145543& 0.096873 &  0.099255& 0.089945& 0.081806& 0.116859 &  0.098330  \\
				l(CI)   &-0.69998 &-0.032129& 0.193988  & 0.790919& 0.148051  &-0.129151&-0.128346& 0.939001& 0.319561 &  0.998701   \\
				CP      &0.983469 & 0.952432& 0.932298  & 0.982363& 0.988270 &  0.930314& 0.973451 &0.899999& 0.948021 &  0.982681 \\
				\hline
				\hline
				&\multicolumn{10}{c|}{Elastic net}\\ \cline{2-11}
				bias    &0.001134 & 0.001182& 0.001025  & 0.001295& 0.001137 & -0.001222& -0.001950  &-0.001144&-0.000927 &-0.000946     \\
				rel.bias&0.113446 & 0.118207& 0.102551  & 0.129495& 0.113726 & -0.122161& -0.195015  &-0.114449&-0.092724 &-0.094431  \\
				SD      &0.233565 & 0.545904& 4.778011  & 3.183053& 0.353661 &  0.758662&  0.294977  & 0.270722& 0.296495 & 0.386398 \\
				SE      & 0.007386 & 0.017263& 0.151094 & 0.100657& 0.011184 &  0.023991&  0.009328  & 0.008561 &0.009376 & 0.012219 \\
				RMSE    & 0.002275 & 0.017697& 0.124326 & 0.116548& 0.095538 &  0.052499&  0.041814  & 0.037475 &0.030259 & 0.030408  \\
				l(CI)   &-0.156817 &-0.078659&-0.028021 & 0.013650& 0.014928 & -0.014636& -0.024378  &-0.110285 &0.120760 & 0.016293   \\
				CP      & 0.983495 & 0.989006& 0.980432 & 0.973206& 0.976353 &  0.951783&  0.964109  & 0.956710 &0.957808 & 0.987693  \\
				\hline
				
			\end{tabular}
		}
		\caption{Simulation results for scenario 2 ($50\%$ zero-inflation proportion) with $n = 1000$}
		\label{t7}
	\end{table}
	
	\begin{table}[h]
		\centering
		\begin{tabular}[h]{|l|c|c|c|}
			\hline
			\backslashbox{Method}{$n$} & 100 & 500&  1000  \\
			\hline 
			\multicolumn{4}{c|}{Scenario 1 ($25\%$ zero inflation)} \\ \cline{2-4}	 
			\hline
			Lasso        &-52.08375 (447.1399) &-295.637 (229.3316)  & -603.0834 (120.936)  \\ 
			Ridge        &-57.7041 (417.4873)  & -300.2647 (229.1527)   & -599.1552 (114.49) \\ 
			Elastic net  &-54.84671 (410.0062)  & -297.7286(215.4572) & -619.984 (106.451)\\ \hline
			\multicolumn{4}{c|}{Scenario 2 ($50\%$ zero inflation)} \\ \cline{1-4}	 
			\hline
			Lasso & -36.96275  (1302.699) &-202.7017 ( 767.2399)  &  -37.23333   (532.699)\\ 
			Ridge &-32.6073 (1195.53578)   &-216.6431  (747.8487)    & -427.7268(541.1651)   \\ 
			Elastic net &-40.24438 (1006.1871)  & -248.5594 (729.9924)&-426.3986 (584.2824) \\ 
			\hline
		\end{tabular}
		\caption{Log-likelihood penalized by $L_1$ (LASSO), $L_2$ (Ridge) and Elastic-net in the two scenarios given with AIC in brackets for $n=1000$.}
		\label{t8}
	\end{table} 

\section{Real data application}
\label{sec:application}

\section{Discussion and perspectives}
\label{sec:disc}
In this paper, we have considered the problem of regularization and estimating in the zero-inflated Bernoulli regression model. The estimator we propose is obtained by maximizing a penalized  likelihood function, which is derived from a joint regression model for the binary response of interest and the zero inflation component considered as a random variable whose distribution is modeled by a logistic regression. We considered three regularization methods: LASSO, ridge and elastic-net.~\\
we have estabilished the asymptotic properties (existence, consistency, and asymptotic normality) of the proposed maximum likelihood estimators using the three methods respectively and investigated their finite-sample properties via simulations.~\\
We then compared the performance of the three methods using simulation results. We found that the elastic-net regularization method provided the best results.\\
In regression analysis for binary data, it is usually of interest to estimate the probability of infection $\pi(\mathbf x)=\mathbb P(Y=1|\mathbf X=\mathbf x)$, for some given value $\mathbf x$ of the covariates and to investigate its properties. Another issue of interest deals with the inference regularization in the zero-inflated Bernoulli regression regression model with a generalized extreme value link function for the the zero inflation component. This situation can be very effective in the case of rare events.

\setlength {\bibsep }{0pt}
\bibliographystyle{Chicago}

\end{document}